\documentclass[global,twocolumn]{svjour}
\usepackage{epsfig}

\usepackage{graphics}
\journalname{Applied Physics B}

\newcommand{\ket}[1]{\ensuremath{\left|{#1}\right\rangle}}

\def\be{\begin{equation}}
\def\ee{\end{equation}}
\def\eea{\end{eqnarray}}
\def\bea{\begin{eqnarray}}

\newcommand{\exs}[1]{\ensuremath{\langle{#1}\rangle}}

\newcommand{\vr}{\ensuremath{\varrho}}

\newcommand{\ketbra}[1]{\ensuremath{| #1 \rangle \langle #1 |}}

\newcommand{\kommentar}[1]{}
\newcommand{\mean}[1]{\ensuremath{\langle{#1}\rangle}}
\newcommand{\PP}{\ensuremath{\mathcal{P}}}

\begin{document}
\title{Detection of multipartite entanglement with two-body correlations}
\author{G\'eza T\'oth\inst{1} \and Otfried G\"uhne \inst{2}
}                     
%
%
\institute{ Max-Planck-Institut f\"ur
Quantenoptik, Hans-Kopfermann-Stra{\ss}e 1, D-85748 Garching,
Germany\and Institut f\"ur Quantenoptik und Quanteninformation,
\"Osterreichische Akademie der Wissenschaften,
A-6020 Innsbruck, Austria}
%
%
\maketitle

\begin{abstract}
We show how to detect entanglement with criteria
built from simple two-body correlation terms.
Since many natural Hamiltonians are sums of
such correlation terms, our ideas can be used to
detect entanglement by energy measurement.
Our criteria can straightforwardly be applied
for detecting different forms of multipartite
entanglement in familiar spin models in thermal
equilibrium.
\end{abstract}

\section{Introduction}
\label{intro}

Entanglement is an important non-classical phenomenon in quantum
mechanics which plays also a crucial role in the novel field of
Quantum Information Theory. While for pure quantum states it is
equivalent to correlations, for mixed states the two notions
differ. In this general case, an $N$-qubit quantum state is
entangled if its density matrix cannot be written as a convex
sum of product states
\begin{equation}
\rho =
\sum_l p_l \rho_l^{(1)} \otimes \rho_l^{(2)}
\otimes ...\otimes \rho_l^{(N)}. \label{sep}
\end{equation}
States of the form Eq.~(\ref{sep}) are called separable.
Based on this definition, several sufficient conditions
for entanglement have been developed. In special cases,
e.g. for $2\times 2$ (two-qubit) and $2\times 3$ (qubit-qutrit)
bipartite systems \cite{AP96,H96AA} and for bipartite multi-mode Gaussian
states \cite{GK01} even necessary and sufficient conditions
are known.

However, in an experimental situation usually only limited
information about the quantum state is available. In this
case,  only those
approaches for entanglement detection can be applied which
require the measurement of not too many observables.
One of such approaches is using entanglement witnesses.
They are entanglement conditions which are linear in
expectation values of observables.
The theory of entanglement witnesses has
recently been rapidly developing \cite{H96AA,H96,LK01,KW04}. It has been shown how
to generate entanglement witnesses that detect states close to a
given one, even if it is mixed or a bound entangled state
\cite{AB01,PH04}. It is also known how to optimize a witness operator
in order to detect the most entangled states \cite{LK00}. Apart
from constructing witnesses, it is also important to find a way to
measure them. Optimal measurement of witnesses have been studied
in Refs. \cite{decomp,stabwit,stabwit2}. Recently, witnesses have
been developed to detect entanglement in physical systems in the
thermodynamical limit \cite{T05,WB04,GT05}.

Entanglement witnesses can not only be used to detect
entanglement experimentally, but can also be used to
characterize the entanglement of a multipartite quantum
state. As we will see later, in the multipartite setting
several different classes of entanglement occur, and entanglement
witnesses can be used to decide in which class a given state is
\cite{AB01}.

In this paper we ask what we can do for systems of very many
particles, e.g., for spin chains in thermal equilibrium.
Entanglement in spin chains has already been extensively studied
\cite{SPINMODELS}. In Section 2 we discuss how to detect
entanglement in general in spin models based on the ideas
presented in Ref. \cite{T05}. In Section 3 we study the detection
of different types of multipartite entanglement in these systems
as discussed in Ref. \cite{GT05}. For this aim, we determine what
the important questions are from this point of view in spin
systems in the thermodynamical limit. Then we look for appropriate
entanglement witnesses, which are easy to construct and study
multipartite entanglement with them \cite{multi}.

\section{Bipartite entanglement}

Let us consider first the two-qubit case.
The simplest expression which can be used
for entanglement detection must contain at
least two correlation terms
\begin{equation}
W:=A^{(1)}A^{(2)}+B^{(1)}B^{(2)},
\end{equation}
where $A_k$ and $B_k$ are operators acting on qubits $k=1,2.$
For simplicity, let us consider $A_k$ and $B_k$ with eigenvalues $\pm 1.$
Now, if we want to use $W$ for entanglement detection,
we have to make sure that
\begin{equation}
\inf_{\Psi} \exs{W}_\Psi < \inf_{\Phi\in \PP}  \exs{W}_\Phi,
\label{larger}
\end{equation}
where the right hand side of the equation is minimized over the
set of product states $\PP$. (Minimization over all mixed
separable states would lead to the same value due to the convexity
of the set of separable states.) Eq.~(\ref{larger}) expresses the
fact that the minimum of $\exs{W}$ must be larger for separable
states than for quantum states in general. For that it is
necessary to have \cite{stabwit2}
\begin{eqnarray}
{[}A_k,B_k]&\ne& 0, \label{comm}
\end{eqnarray}
for $k=1,2.$ Here $[..]$ denotes the commutator. Eq.~(\ref{comm})
expresses the fact that we have to measure two different
observables at each party. For entanglement detection in an experiment,
the ratio of the two minima in
Eq.~(\ref{larger}) must be the largest possible. It is
straightforward to see that this is the case if we choose
operators such that
\begin{eqnarray}
\{A_k,B_k\}&=& 0,
\end{eqnarray}
for $k=1,2.$ Here $\{..\}$ denotes the anticommutator. An example
for such an operator is then
\begin{equation}
h_{XY}:=X^{(1)}X^{(2)}+Y^{(1)}Y^{(2)},
\label{hXY}
\end{equation}
where $X$ and $Y$ denote Pauli spin matrices.
For this operator the minimum of the expectation value is
\begin{eqnarray}
\inf_{\Psi} \exs{h_{XY}}_\Psi&=&-2.
\end{eqnarray}
The state giving the minimum is the two-qubit singlet
\begin{equation}
\ket{\psi_s}:=\frac{1}{\sqrt{2}}(\ket{10}-\ket{01}). \label{singlet}
\end{equation}
For this state $\exs{X^{(1)}X^{(2)}}=\exs{Y^{(1)}Y^{(2)}}=-1.$
The minimum
for product states can be obtained as follows. For product states
we have
\begin{eqnarray}
\exs{h_{XY}}=\exs{X^{(1)}}\exs{X^{(2)}}+ \exs{Y^{(1)}}\exs{Y^{(2)}} \ge - 1,
\end{eqnarray}
where the last inequality follows from the Cauchy-Schwarz inequality and
knowing that $\exs{X^{(k)}}^2+\exs{Y^{(k)}}^2\le 1.$
Among operators with three correlation terms used for entanglement detection
the following form is optimal
\begin{equation}
h_{H}:=X^{(1)}X^{(2)}+Y^{(1)}Y^{(2)}+Z^{(1)}Z^{(2)}.
\label{hH}
\end{equation}
The minimum of the expectation value of $h_{H}$ is $-3$.
For separable states the minimum is $-1$
which can be proved similarly as it has been done for $h_{XY}.$

Now let us move to the $N$-qubit case. Consider the expression
\begin{equation}
H_{XY}:=J \sum_{k=1}^N X^{(k)}X^{(k+1)}+Y^{(k)}Y^{(k+1)},
\label{HXY}
\end{equation}
where $J>0$ is the coupling constant and according to the usual
assumption for a periodic boundary condition qubit $(N+1)$ is
identical to qubit $(1).$ This is the Hamiltonian for the XY
chain. The minimum for separable states is now
\begin{eqnarray}
\inf_{\Phi\in \PP} \exs{H_{XY}}_\Phi&=&-JN.
\end{eqnarray}
This comes from knowing that for product states each term in the
summation in Eq.~(\ref{HXY}) is bounded by $-1$ as we have seen it
before. The minimum for quantum states can be obtained from
numerical calculations since the $XY$ model is solvable \cite{XYmodel}.
Similarly, we can define
\begin{equation}
H_H:=J \sum_{k=1}^N
X^{(k)}X^{(k+1)}+Y^{(k)}Y^{(k+1)}+Z^{(k)}Z^{(k+1)}, \label{HH}
\end{equation}
This is the Hamiltonian for the Heisenberg chain.
The minimum for
separable states is
\begin{eqnarray}
\inf_{\Phi\in \PP} \exs{H_H}_\Phi&=&-JN.
\end{eqnarray}
The minimum for quantum states can be obtained for large $N$ as
\cite{T99}
\begin{equation}
\inf_{\Psi} \exs{H_H}_\Psi=-4\bigg(\ln 2-\frac{1}{4}\bigg)NJ \approx -1.77NJ.
\end{equation}

For our spin chain Hamiltonians
the ratio between the minimum for general quantum states and the minimum for separable states
is smaller than for Eqs.~(\ref{hXY},\ref{hH}) since
there is not a quantum state saturating all two-body correlation terms.
In fact, it is easy to see that there is not a Hamiltonian built from two-body correlations such that its unique ground state saturates all correlation terms and this ground state is true multipartite entangled \cite{STAB}.

What are the advantages of the expressions Eq.~(\ref{HXY}) and
Eqs.~(\ref{HH}) in detecting entanglement? They are easily
measurable locally, since they are the sum of only a few two-body
correlation terms. Moreover, in some physical systems
Eq.~(\ref{HH}) can directly be measured as the average
nearest-neighbor correlation, or as the energy of the system if
this system can be described by a Heisenberg Hamiltonian.

\begin{figure}
\centerline{\epsfxsize=3.0in\epsffile{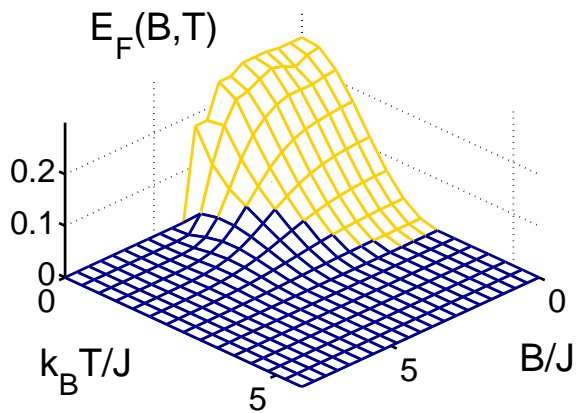}}
\caption{Heisenberg chain of $10$ spins.
Nearest-neighbor entanglement as a function of magnetic field $B$ and
temperature $T$.}
\label{fig_Heisenberg}
\end{figure}

The previous ideas can straightforwardly be applied to spin chains
in thermal equilibrium \cite{T05}. Let us consider the Heisenberg
Hamiltonian in an external magnetic field $H_{HB}:=H_H+B\sum_k Z^{(k)}.$
For this Hamiltonian, it is easy to bound the minimum for separable
states \cite{T05}. Any time $\exs{H_H}$ is below this value, we know
that the thermal state is entangled. This is demonstrated in
Fig.~\ref{fig_Heisenberg}. It shows the nearest-neighbor entanglement
vs. $B$ and the temperature $T$. The entanglement of formation was
computed from the concurrence \cite{BV96}. Light color indicates the
region where the thermal ground state is detected as entangled
based on the ideas discussed before. As one can see, there are regions
with $E_F>0$ which are not detected. However, it can be seen in the Figure
that when the system contains at least a small amount of entanglement
($\sim 0.07$) the state is detected as entangled. Note that the sharp
decrease of the nearest-neighbor entanglement around $B_{crit}=4$ for
$T=0$ is due to a {quantum phase transition}.

\section{Multipartite entanglement}

In this section we will discuss how to detect multi-party
entanglement by measuring the operators described before.
Our motivation is that entanglement for many particles
is qualitatively different from the two-party case, and
many new phenomena arise \cite{pheno}.
There are several possibilities to classify entanglement
of many parties. We are now looking for the terminology
which is appropriate for spin chains of many particles.

Let us first recall the notion of genuine multipartite
entanglement. A pure state $\ket{\psi}$ of a
quantum system of $N$ parties is called {\it fully separable}
if it is a product state for all parties,
$\ket{\psi} = \ket{\phi_1} \otimes\ket{\phi_2}\otimes ...\otimes \ket{\phi_N}.$
It is  called {\it biseparable}, when a partition of the
$N$ parties into two groups $A$ and $B$ can be found,
such that the state is a product state with respect
to this partition, namely
\be
\ket{\psi}=\ket{\phi_A} \otimes \ket{\phi_B}.
\ee
If this is not the case, the state is called {\it genuine
multipartite entangled.} Note that the vectors $\ket{\phi_A}$
and $\ket{\phi_B}$ are allowed to contain entanglement within
their partition. Thus, to prove genuine multipartite entanglement,
it does not suffice to exclude full separability.

For mixed states, these definitions can, as usual, be extended
via convex combinations. Indeed, the definition of full separability
was already given in Eq.~(\ref{sep}). A mixed state is biseparable,
whenever we can write $\vr=\sum_i p_i \ketbra{\psi_i}$ with
biseparable $\ket{\psi_i}$ and some probabilities $p_i.$ Here,
the states $\ket{\psi_i}$ are allowed to be biseparable with
respect to different partitions.

Another approach to classify multipartite entanglement asks
whether multipartite entanglement is necessary to form a
given state \cite{GT05}. In this approach, a state $\ket{\psi}$
{\it producible by $k$-party entanglement} (or $k$-producible, in short)
if we can write the state $\ket{\psi}$ as a tensor product
\be
\ket{\psi}=\ket{\phi_1}\otimes\ket{\phi_2}\otimes
...\otimes \ket{\phi_m},
\ee
where the states $\ket{\phi_i}$ are states on maximally $k$-qubits.
In this definition, a two-producible state does not contain any
multipartite entanglement, since it suffices to generate the two-qubit
states $\ket{\phi_i}$ to arrive at the state $\ket{\psi}.$
In addition, we say that a state {\it contains genuine
$k$-party entanglement} if it is not producible by $(k-1)$-party
entanglement.
This definition can be extended to mixed states as before via convex
combinations. Again, a mixed state which is $k$-producible requires
only the generation of $k$-party pure entangled states and mixing
for its production (see also Fig.~2). Consequently, {a mixed state
$\vr$ contains $k$-party entanglement, iff the correlations
cannot be explained by assuming the presence of $(k-1)$-party
entanglement only in the pure subsensembles.}

The notions of genuine multipartite entanglement and producibility
are not completely independent. For example, the states containing
$N$-party entanglement are just the genuine multipartite entangled
states and the one-producible states are fully separable. If one
can show that a reduced state of $k+1$ qubits is genuine multipartite
entangled, then this implies that the total state is not $k$-producible,
while the converse is in general not true.

For spin chains of {\it macroscopic} size, it is in general very
difficult to prove that the total state is genuine $N$-partite
entangled via energy measurements. This is due to the fact that
the notion of genuine $N$-partite
entanglement is extremely sensitive to the properties
of a single qubit. Indeed, in order to prove genuine multipartite
entanglement, one has to exclude the possibility, that one single
qubit can be separated from the remaining $N-1$ qubits. However,
multipartite entanglement in the {\it reduced} states of small numbers
of qubits can easily be detected, as we will see.
Moreover, if the reduced state is multipartite entangled then
the state is not two-producible.

 \begin{figure}
\centerline{\epsfxsize=3.2in\epsffile{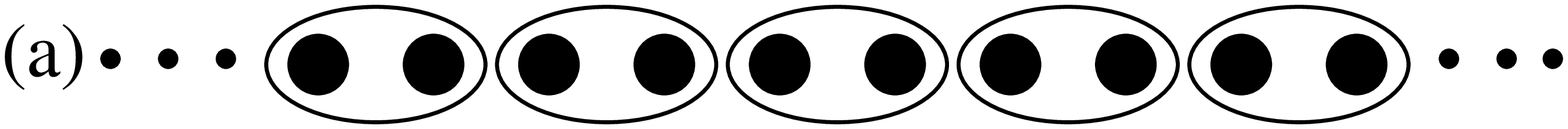}}
\centerline{\epsfxsize=3.2in\epsffile{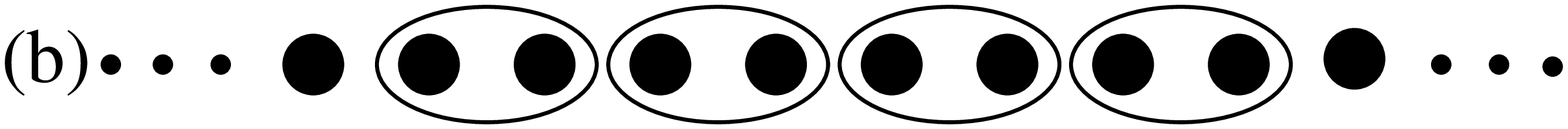}}
\caption{(a) Chain of two-qubit singlets. (b) The same state shifted
by one qubit to the right. The mixture of these two states is two-producible,
 that is, does not need three-qubit entanglement when produced from
pure states by mixing.
}
\label{fig_singlet}
\end{figure}

Now let us see our results for the XY model and the Heisenberg chain.
The proofs for the following theorems are given in Ref.~\cite{GT05}.
We always assume periodic boundary conditions and that the number
of spins $N$ is even.

{\bf Theorem 1.}
Let  $\vr$  be an $N$ qubit state whose dynamics
is governed by the XY-Hamiltonian in Eq.~(\ref{HXY}).
If $\vr$ is one-producible (fully separable), then
\be
{\mean{H_{XY}}} \geq  -J N
\label{XXmodel1}
\ee
holds. If ${\mean{H_{XY}}} < -J N$ this implies that
there are two neighboring qubits such that their
reduced state is entangled.
For two-producible states
\be
{\mean{H_{XY}}} \geq  - \frac{9}{8}J N
\label{XXmodel2}
\ee
holds. If ${\mean{H_{XY}}} < - {9}/{8}J N$ the state contains thus
tripartite entanglement and if
\be
{\mean{H_{XY}}} < - \frac{1+\sqrt{2}}{2} J N  \approx - 1.207 J N
\label{XXmodel3}
\ee
then there exist three neighboring qubits $i, i+1, i+2$
such that the reduced state $\vr_{i,i+1,i+2}$
of these qubits is genuine tripartite entangled.

For the Heisenberg model, we can state the following:

{\bf Theorem 2.} Let  $\vr$  be an $N$ qubit state with the Heisenberg
Hamiltonian of Eq.~(\ref{HXY}).
If $\vr$ is one-producible (fully separable), then
\be
{\mean{H_H}} \geq  -J N
\label{XXXmodel2}
\ee
holds, while for two-producible states
\be
{\mean{H_H}} \geq  - \frac{3}{2}J N
\label{XXXmodel3}
\ee
holds. Thus, if ${\mean{H_H}} < 3N/2$ the state contains
genuine tripartite entanglement.
Furthermore, if
\be
{\mean{H_H}} < - \frac{1+\sqrt{5}}{2}JN \approx - 1.618 J N
\label{XXXmodel1b}
\ee
then there are three neighboring qubits such that their
reduced state is genuine tripartite entangled.

Let us see an example. Consider the state shown in
Figure~\ref{fig_singlet}(a)
\begin{equation}
\ket{\Phi_{s}}=\ket{\psi_s} \otimes \ket{\psi_s}\otimes \ket{\psi_s}\otimes ... ,
\end{equation}
where the two-qubit singlet is defined in Eq.~(\ref{singlet}). It is easy to see that state $\ket{\Phi_{s}}$ saturates the inequality
Eq.~(\ref{XXXmodel3}). It is not surprising, since it is a two-producible state.
Let us now define operator $S$ which shifts the qubits
by one, i.e.,
\begin{eqnarray}
&& S \ket{\alpha_1} \otimes \ket{\alpha_2}\otimes \ket{\alpha_2}\otimes ...  \otimes \ket{\alpha_N}\nonumber\\
&& \;\;\;\;\;\;\;\; = \ket{\alpha_N} \otimes \ket{\alpha_1}\otimes \ket{\alpha_2}\otimes ...  \otimes \ket{\alpha_{N-1}}.
\end{eqnarray}
Consider the state
\begin{equation}
\rho_m:=\frac{1}{2}\bigg(\ketbra{\Phi_s}+S\ketbra{\Phi_s}S^\dagger \bigg).
\end{equation}
This state is the mixture of the singlet chains depicted in
Figure~\ref{fig_singlet}(a) and (b). $\rho_m$ is not fully
separable and is not a product of single-qubit and two-qubit
density matrices.  Moreover, the state $\rho_m$ has a negative
partial transpose with respect to each partition.
However, $\rho_m$ also saturates the inequality Eq.~(\ref{XXXmodel3})
and it is also two-producible. That is, three-qubit entanglement
is not needed to create it, and it contains no multipartite
entanglement.

The previous results can straightforwardly be used for obtaining a limit
temperature for the different forms of entanglement. We define thus the
temperatures $T_{R2},$ $T_{R3},$ $T_{C2}$ and $T_{C3}$ below which either
reduced states of two or three parties are entangled or the
total state contains two or three-party entanglement.
Obviously, $T_{R2}= T_{C2} > T_{C3} > T_{R3}$
has to hold here. These temperature bounds are shown for a
Heisenberg chains of a couple of spins in Table 1. As expected,
the values for $ T_{C2}=T_{R2}$ coincide with the ones of Ref.~\cite{T05}.
The given values for $T_{C3}$ and $T_{R3}$ show that in the Heisenberg
chain of ten spins at $k_BT \approx J$ multipartite entanglement plays a
role, namely at least one reduced state is genuine tripartite
entangled and the total state contains tripartite entanglement.

\begin{table}
\caption{
Threshold temperatures $T_{C2}, T_{R3}$ and $T_{C3}$
for a Heisenberg chain. The parameters as set to
$J=k_B=1.$ See text for details.}

\begin{center}
\begin{tabular}{|c||c|c|c|c|c|}
\hline
 $N$  & 2 & 4 & 6 & 8 & 10 
\\
\hline
\hline
$T_{C2}$&   7.28 &    3.45      &3.21  &3.18  & 3.18 
\\
\hline
$T_{C3}$&  - \kommentar{4.9702} &2.10&1.75& 1.65 & 1.62
\\
\hline
$T_{R3}$&  - \kommentar{4.6040}& 1.85&1.46& 1.32 &1.26
\\
\hline
\end{tabular}
\end{center}
\end{table}

\section{Conclusions}

We discussed how to construct entanglement
conditions using two-body correlations. This
implies, that typical Hamiltonians as appearing
in the XY model or the Heisenberg model can serve
for entanglement detection in spin models.
Also different forms of multipartite entanglement
can be detected in this way.

A natural continuation of our work lies in the
extension of our bounds to other systems. Here,
spin systems in two or three dimensions as well
as frustrated systems are of interest. Furthermore,
it would be also desirable to derive energy bounds
also for higher classes of multipartite entanglement,
e.g. three-producible states.

We would like to thank H.J. Briegel, J.I. Cirac, A.C. Doherty, and
M. Dowling for discussions. We acknowledge the support of the
European Union (Grant No. MEIF-CT-2003-500183), the EU projects
QUPRODIS, RESQ, SCALA and OLAQUI, the DFG and the
Kompetenznetzwerk Quanteninformationsverarbeitung der Bayerischen
Staatsregierung.

%
%

\end{document}